# Colossal magnetoresistance in a Mott insulator via magnetic field-driven insulator-metal transition

M. Zhu[1], J. Peng[2], T. Zou[1], K. Prokes[3], S. D. Mahanti[1], T. Hong[4], Z. Q. Mao[2], G. Q. Liu[5], X. Ke[1]*

[1] *Department of Physics and Astronomy, Michigan State University, East Lansing, Michigan 48824, USA*

[2] *Department of Physics and Engineering Physics, Tulane University, New Orleans, Louisiana 70118, USA*

[3] *Helmholtz Zentrum Berlin, D-14109 Berlin, Germany*

[4] *Quantum Condensed Matter Division, Oak Ridge National Laboratory, Oak Ridge, Tennessee 37831, USA*

[5] *Ningbo Institute of Material Technology and Engineering, Chinese Academy of Sciences, Ningbo 315201, China*

*Corresponding author: ke@pa.msu.edu.

We present a new type of colossal magnetoresistance (CMR) arising from an anomalous collapse of the Mott insulating state via a modest magnetic field in a bilayer ruthenate, Ti-doped $Ca_3Ru_2O_7$. Such an insulator-metal transition is accompanied by changes in both lattice and magnetic structures. Our findings have important implications because a magnetic field usually stabilizes the insulating ground state in a Mott-Hubbard system, thus calling for a deeper theoretical study to reexamine the magnetic field tuning of Mott systems with magnetic and electronic instabilities and spin-lattice-charge coupling. This study further provides a model approach to search for CMR systems other than manganites, such as Mott insulators in the vicinity of the boundary between competing phases.

Understanding and controlling the complex behaviors of strongly correlated electron systems is a central topic of condensed matter research. The strong interplay between electron charge, spin, and orbital degrees of freedom renders collective behaviors in their physical properties which are extremely sensitive to external perturbations such as chemical doping, magnetic field, and pressure. Colossal magnetoresistance (CMR), a change in resistivity by orders of magnitude driven by magnetic field, is one of the exotic phenomena seen in strongly correlated electron systems, which has promising applications in information storage devices. Prototypical materials displaying CMR are mixed-valence perovskite manganese oxides. The itinerant $e_g$ electrons mediate ferromagnetic coupling between localized $t_{2g}$ moments through the double exchange mechanism leading to a ferromagnetic metallic state. The competition between ferromagnetic double exchange and antiferromagnetic superexchange, together with Jahn-Teller distortion which gives rise to charge-orbital ordering and polaronic transport, constitutes the basis of the current understanding of the CMR physics in doped manganites [1, 2].

A Mott insulator has an insulating ground state driven by Coulomb repulsion. Generally, there are two fundamental tuning parameters in a Mott system, i.e., bandwidth and band filling [3], which can be controlled by varying pressure, carrier doping, electric field, etc [4-7]. However, a collapse of the Mott gap, i.e., insulator-to-metal transition (IMT), induced by a magnetic field is rarely observed. Theoretical studies based on a half-filled single-band Hubbard model predicted a magnetic field-driven first-order metamagnetic localization transition from a strongly correlated metallic paramagnetic state to an insulating state via a competition between the local antiferromagnetic exchange and the Zeeman energy [8-10]. This prediction has been suggested to account for the field-induced Mott transition in a quasi-two-dimensional organic conductor κ-(BEDT-TTF)$_2$Cu[N(CN)$_2$]Cl [11]. However, Mott insulators are often antiferromagnetic, and the metal-insulator transition (MIT) is accompanied by lattice distortion. The effect of strong coupling between lattice, spin, and orbital degrees of freedom on the electronic structure is crucial to the understanding of the nature of the ground states and associated phase transitions. Hence, the observation of a collapse of a Mott insulating state induced by a magnetic field has significant implications: It not only may give rise to a CMR effect providing a route to search for new CMR materials, but also points to the fundamental role of the coupling among various degrees of freedom and how they collectively respond to a magnetic field, which has not been explored theoretically in great depth.

In this paper we report a new type of CMR phenomenon in a Mott system, Ti-doped $Ca_3Ru_2O_7$ bilayer ruthenate, which occurs concurrently with a magnetic phase transition and a drastic change in the lattice structure. We argue that the underlying mechanism of CMR phenomena observed in Ti-doped $Ca_3Ru_2O_7$ is fundamentally different from that in manganites. Instead, we ascribe it to a magnetic field-driven collapse of the low-temperature Mott insulating ground state.

Ruddlesden-Popper series perovskite ruthentates $(Sr,Ca)_{n+1}Ru_nO_{3n+1}$ are known to exhibit a wealth of fascinating physical properties ranging from unconventional superconductivity to quantum criticality to Mott physics [12-14]. The bilayer ruthenate, $Ca_3Ru_2O_7$, undergoes an antiferromagnetic transition at Néel temperature $T_N \sim 56$ K followed by a metal-insulator transition at $T_{MIT} \sim 48$ K [15]. The metallicity of $Ca_3Ru_2O_7$ reappears below 30 K [16], which originates from a very small (~0.36%), ungapped section of the Fermi surface surviving through the MIT [17]. The magnetic structures below and above $T_{MIT}$ are characterized as ferromagnetic bilayers stacked antiparallel along the $c$ axis [18-20], with the spin direction along the $a$ axis for $T_{MIT} < T < T_N$ (denoted as AFM-a) and along the $b$ axis for $T < T_{MIT}$ (AFM-b).

In this system, the Mott insulating state can be achieved by doping as low as ~3% of Ti into the Ru sites [21]. The Mott character of the charge gap has been confirmed by the recent photoemission measurements [22]. Concomitantly, the magnetic structure transforms from AFM-b to G-type with nearest-neighbor spins coupled antiferromagnetically (G-AFM). The major effect of Ti substitution is to disrupt carrier hopping and narrow the bandwidth [21]. The material of interest in this paper, $Ca_3(Ru_{0.97}Ti_{0.03})_2O_7$, shows a metallic state with the AFM-a type magnetic order for $T_{MIT}$ (~46 K) $< T < T_N$ (~62 K) but a Mott insulating state with the G-AFM order for $T < T_{MIT}$. According to the $T$-$x$ phase diagram of $Ca_3(Ru_{1-x}Ti_x)_2O_7$ (Figure S1 [24]), $Ca_3(Ru_{0.97}Ti_{0.03})_2O_7$ sits right at the boundary between the correlated metallic phase and the Mott insulating phase [23], and the physical properties and responses to external stimuli are to a great extent determined by the competition between these two distinct magnetic and electronic instabilities. Thus, it serves as a model system to explore exotic phenomena of strongly correlated electrons in the vicinity of the MIT.

Figure 1(a) and 1(b) show the temperature dependence of the out-of-plane resistivity $\rho_c$ and in-plane resistivity $\rho_{ab}$ of $Ca_3(Ru_{0.97}Ti_{0.03})_2O_7$ measured with a magnetic field $B$ of 0 T and 9 T applied along the $b$ axis. At zero field, $Ca_3(Ru_{0.97}Ti_{0.03})_2O_7$ undergoes a MIT with decreasing temperature, with a sharp increase in $\rho_c$ and $\rho_{ab}$ by 5 and 3 orders of magnitude, respectively, at the onset of MIT. In contrast, at $B = 9$ T, both $\rho_c$ and $\rho_{ab}$ gradually increase in resistance at low temperature by only a small factor, attributable to disorder scattering. We denote this state as a weakly localized state. Such a dramatic magnetoresistive effect is also observed in isothermal resistivity measurements at $T = 10$ K shown in Figure 1(c) and 1 (d). With increasing $B$ applied along the $b$ axis, both $\rho_c$ and $\rho_{ab}$ decrease sharply at a critical field $B_c \approx 8.5$ T and show a hysteresis with decreasing $B$, indicating a first-order IMT. On the other hand, both resistivity values decrease slightly with the field applied along the $a$ axis which is close to the magnetic hard axis.

To obtain a deeper insight into the magnetic field-induced IMT in $Ca_3(Ru_{0.97}Ti_{0.03})_2O_7$, we performed single crystal neutron diffraction measurements at $T = 10$ K with the field applied along the $b$ axis [24]. Interestingly, the lattice parameters of $Ca_3(Ru_{0.97}Ti_{0.03})_2O_7$ change drastically when the magnetic field approaches the critical value where the resistivity drops significantly. As shown in Figure 2(a), $c$ increases by ~ 0.78%. Similar behavior is observed by heating the sample at zero magnetic field from low temperature to above $T_{MIT}$ (Figure S2 [24]) [21]. These observations clearly demonstrate that the Mott insulating ground state of $Ca_3(Ru_{0.97}Ti_{0.03})_2O_7$ is stabilized by coupling strongly to lattice distortions, similar to the bandwidth-controlled Mott system $(Ca_{1-x}Sr_x)_2RuO_4$ [14, 27]. In both cases, the appearance of the Mott insulating state is accompanied by a structural transition from a long $c$ axis to a short one [21, 28]. The enhanced $RuO_6$ octahedral flattening and tilting below $T_{MIT}$ reduce the bandwidth [33] and change the occupancy of the $t_{2g}$ orbitals [29, 30], leading to a MIT [31-33].

The remarkable lattice structure and resistivity changes induced by an applied magnetic field are accompanied by a magnetic structure transition. Figure 2(c) and 2(d) show the rocking curves of (1 0 2) and (0 0 1) magnetic Bragg peaks, respectively, measured at $T = 10$ K with $B = 0$ T and 10 T after the sample was cooled down in a zero field. Note that the (1 0 2) magnetic Bragg peak refers to the G-AFM magnetic structure in the Ti-doped $Ca_3Ru_2O_7$ [21] while the (0 0 1) Bragg peak corresponds to the AFM-b or AFM-a type magnetic structures in pristine $Ca_3Ru_2O_7$ [20]. The disappearance of the (1 0 2) Bragg peak and the emergence of the (0 0 1) Bragg peak at $B = 10$ T indicate a field-induced magnetic structure change from G-AFM to one similar to AFM-a (or AFM-b). However, the field-induced phase is not a collinear magnetic structure, but a canted antiferromagnetic one (denoted as CAFM), i.e., a vector superposition of an AFM-a type structure and a ferromagnetic component along the $b$ axis [Figure 3(b), see supplemental materials for details]. The strong coupling between the field-induced structural transition and the magnetic transition can be readily seen. The integrated intensity of (1 0 2) and (0 0 1) Bragg peaks measured at 10 K as a function of increasing $B$ is shown in Figure 2(b). The (0 0 1) Bragg peak intensity increases significantly at $B_c$ where the (1 0 2) diffraction intensity has a sharp drop, indicative of a magnetic phase transition. Clearly, the critical field of the magnetic transition coincides with that of the structural transition. These observations are similar to those observed in the $T$-dependent studies in the same compound but in the absence of a magnetic field, reinforcing the presence of a strong spin-lattice-charge coupling in this system [21]. The (1 0 2) and (0 0 1) magnetic Bragg peak intensities as a function of $B$ at some representative temperatures are shown in Figure S5 [24]. The hysteresis observed at low temperature indicates the first-order nature of the field-induced transition between G-AFM and CAFM. The CMR effect can be observed only accompanying the field-driven G-AFM to CAFM transition below $T_{MIT}$. On the contrary, the resistivity shows little change for $T_{MIT} < T < T_N$, where (0 0 1) is suppressed continuously as the field increases, until the magnetic moments are fully polarized by the magnetic field.

Figure 3(a) presents the $B$-$T$ phase diagram which summarizes our electronic transport and neutron diffraction measurements. At $T_{MIT} < T < T_N$, upon applying the magnetic field along the $b$-axis $Ca_3(Ru_{0.97}Ti_{0.03})_2O_7$ undergoes magnetic phase transitions from an AFM-a to a CAFM structure [Figure 3(b)] and finally to a fully polarized paramagnetic (PM) state with further increasing the field, a feature similar to that observed in the parent compound [20]. In contrast, below $T_{MIT}$ $Ca_3(Ru_{0.97}Ti_{0.03})_2O_7$ has a magnetic phase transition from G-AFM to CAFM, which occurs simultaneously with the electronic structure changed from an insulating state to a weakly localized state, implying the strong correlation between the field-induced IMT and magnetic phase transition.

We would like to point out that the physical origin of the CMR effect observed in the Ti-doped $Ca_3Ru_2O_7$ system is fundamentally distinct from that in the hole-doped manganites. The canonical CMR effect is observed in manganites that become ferromagnetic via the double-exchange mechanism [34] at low temperature, such as $La_{1-x}Sr_xMnO_3$ ($x = 0.175$) [2]. An application of a magnetic field leads to a decrease in electrical resistance with large magnetoresistance seen only near the transition temperature. This phenomenon is ascribed to reduction in spin scattering and polaron effect due to strong electron-phonon interaction. The occurrence of CMR can also originate from the field-induced melting of the charge-ordered state in manganites, for example $Pr_{0.5}Ca_{0.5}MnO_3$ and $Nd_{0.5}Sr_{0.5}MnO_3$, with a ratio of $Mn^{3+}$ to $Mn^{4+}$ being commensurate with the crystal lattice [35, 36]. However, neither of these two mechanisms can be applied to account for the magnetoresistance observed in $Ca_3(Ru_{0.97}Ti_{0.03})_2O_7$ considering the fact that the CMR effect emerges only at low temperature (below $T_{MIT}$) and the single-valence state of Ru ions ($Ru^{4+}$). Furthermore, the CMR effect in manganites can be explained using the one-$e_g$ orbital model [37], although a later study using a two-$e_g$ orbital model explained the first-order CMR transition for some manganites [38]. In contrast, it is essential to take into account the multiband ($t_{2g}$ orbitals) structure of ruthenates in order to understand the MIT (see more discussion later).

The magnetoresistive effect in $Ca_3(Ru_{0.97}Ti_{0.03})_2O_7$ is also distinct from that in the parent compound. (1) Importantly, the parent $Ca_3Ru_2O_7$ compound exhibits a metallic ground state; thus, no magnetic field-induced electronic phase transition occurs. (2) When the magnetic field is applied along the easy axis (*b* axis), an enhanced conductivity by ~1 order has been observed which is accompanied by a first-order metamagnetic transition. Based on neutron diffraction measurements, the large magnetoresistance in $Ca_3Ru_2O_7$ was ascribed to a bulk spin-valve effect associated with spin scattering [20]. (3) Although a reduction of resistivity by 3 orders of magnitude in the metallic ground was observed with a field of 15 T applied along the magnetic hard axis (*a* axis) at $T = 0.4$ K, which was argued to originate from the suppression of orbital ordering at high fields [39], no orbital ordering in $Ca_3Ru_2O_7$ at zero field was convincingly detected in recent resonant *x*-ray diffraction experiments [40]. In contrast to the case of the parent compound, the first-order magnetoresistance transition observed in $Ca_3(Ru_{0.97}Ti_{0.03})_2O_7$ [Figure 1(a) and 1(b)] cannot be interpreted in the framework of a spin-valve mechanism; instead, it suggests a magnetic field-induced change of the electronic ground state from a Mott insulator to a weakly localized state, i.e., a collapse of the Mott state.

Within a quasiparticle picture, the closure of the Mott gap by a modest magnetic field ($B_c \approx$ 8.5 T at 10 K) seems unlikely, considering that the upper limit of the associated Zeeman energy/$Ru^{4+}$ spin is about 1.4 meV, much smaller than the Mott gap of ~120 meV estimated from optical conductivity measurements below $T_{MIT}$ at zero field in $Ca_3(Ru_{0.97}Ti_{0.03})_2O_7$ [41]. Therefore, other factors must be taken into account to explain this transition. In particular, our experimental results indicate several key aspects which are crucial to the understanding of the field-induced IMT in $Ca_3(Ru_{0.97}Ti_{0.03})_2O_7$.

First of all, due to the proximity to the phase boundary (Figure S1 [24]), two dramatically different states, one a weakly localized state with ferromagnetic bilayer spin structure (AFM-a/AFM-b) and the other a Mott insulating state with a G-AFM type magnetic structure, are close in energy. The significance of the delicate competition between different phases, which is also believed to give rise to the CMR effects in manganites [42, 37], is manifested by the fact that for materials slightly away from the phase boundary, for example $x = 0.05$, no magnetic field-induced IMT [24] is observed (Figure S6 [24]) up to 9 T.

Second, electron-lattice coupling plays an essential role here. As discussed above, in $Ca_3(Ru_{0.97}Ti_{0.03})_2O_7$, at zero field and below $T_{MIT}$ the G-AFM phase with antiferromagnetic intrabilayer coupling is associated with a short $c$-axis lattice structure while above $T_{MIT}$ the AFM-a phase with ferromagnetic intrabilayer interaction prefers a long $c$-axis structure (see Figure S2 [24]). The lattice distortions have a dramatic effect on magnetism and electronic structure in the three-band Hubbard system [33, 43], such as the single-layer $Ca_{2-x}Sr_xRuO_4$ system [29, 30]. It was shown that the rotation of the $RuO_6$ octahedron enhances ferromagnetism and subsequent tilting favors antiferromagnetism while the flattening of $RuO_6$ favors both [24]. All these effects compete and give rise to a strong correlation between the lattice and magnetic structures in $Ca_{2-x}Sr_xRuO_4$. For Ti-doped $Ca_3Ru_2O_7$, a similar picture also holds, as suggested by the projected density of states (Figure 4) obtained using density functional theory (DFT) calculations [24]. When the magnetic field is applied along the $b$ axis below $T_{MIT}$, the G-AFM spin configuration becomes unstable due to the competition among Zeeman energy, magnetocrystalline anisotropy, and exchange interaction. As a result, above a critical field a new magnetic ground state (CAFM) with ferromagnetic intrabilayer coupling is favored, which is accompanied by a structural transition with a longer $c$ axis due to the strong spin-lattice coupling.

Last but not least, our DFT calculations suggest that field-induced orbital depolarization in this multiband system may play an important role in the collapse of the Mott insulating ground state in this system [24]. As shown in Figure 4(b), in the spin-down channel the $xy$ orbital is almost fully occupied while the $xz/yz$ orbitals are nearly empty, which suggests an electronic configuration of $xy$ (↑↓) $xz/yz$ (↑,↑) for the insulating phase. This orbital polarization is due to a larger octahedral flattening which lifts the degeneracy of *xy* and *xz* / *yz* $t_{2g}$ orbitals. In contrast, in the metallic state shown in Figure 4(c), the reduced structural distortion gives rise to a suppression of the *xy* orbital occupancy and a drastic reduction of the orbital polarization. The precise orbital configuration of this system is of great current interest and warrants further experimental studies.

In summary, the electronic and magnetic instabilities, in combination with strong interplay of spin-lattice-charge couplings and potential change in the orbital occupancy, enable a drastic change in the electronic structure leading to the collapse of a Mott insulating ground state by a modest applied magnetic field. The above arguments suggest that $Ca_3(Ru_{0.97}Ti_{0.03})_2O_7$ serves as a unique example of using a magnetic field as a controlling parameter to tune the low temperature electronic ground state of antiferromagnetic Mott insulators. We believe that this study will stimulate future theoretical studies on field-induced insulator-metal transition for Mott systems with multiband electronic structure, strong magnetic and electronic instabilities, as well as subtle interplay of various degrees of freedom.

X. K. is grateful for Dr. Matas in HZB for help during the neutron experiment. X. K. acknowledges the start-up funds from Michigan State University. Work at Tulane University was supported by the NSF under Grant No. DMR-1205469 and work at ORNL was supported by the Scientific User Facilities Division, Office of Basic Energy Sciences, DOE. G. L. was supported by the National Natural Science Foundation of China (Grants No. 11204326 and No. 11474296).

**Figure captions**

Fig. 1. Temperature and magnetic field dependence of the resistivity. (a),(b) Temperature dependence of the resistivity, $\rho_c$ and $\rho_{ab}$, measured at $B = 0$ T and 9 T applied along the $b$ axis. The magnetic field was applied at $T = 100$ K and the measurements were taken while cooling. (c),(d) Magnetic field dependence of $\rho_c$ and $\rho_{ab}$ at $T = 10$ K.

Fig. 2. (a) Field dependence of lattice parameters. Statistical errors from the fitting procedure are smaller than the symbol size. (b) Field dependence of the integrated intensity of magnetic Bragg peaks (1 0 2) and (0 0 1), respectively, measured at $T = 10$ K. The magnetic field was increased for both nuclear and magnetic Bragg peak scans. (c),(d) Rocking curves of the (1 0 2) and (0 0 1) magnetic Bragg peaks measured at $B = 0$ T and 10 T applied along the $b$ axis. Measurement temperature $T = 10$ K.

Fig. 3. Phase diagram and the schematics of the magnetic structures. (a) $T$-$B$ phase diagram of $Ca_3(Ru_{0.97}Ti_{0.03})_2O_7$ in a magnetic field along the $b$-axis. The solid squares, circles and diamonds are phase boundaries determined by neutron diffraction measurements: solid squares denote the points where (001) magnetic Bragg peak shows up, and solid circles represent where (102) magnetic Bragg peak disappears. The solid diamonds stand for the points where (001) signal disappears completely. The open diamonds are the phase boundaries determined by the resistivity measurements. (b) Schematics of spin structures of AFM-a, G-AFM, and CAFM phases.

Fig. 4. Projected density of states (PDOS) of the Ru $d_{xy}$ and $d_{xz}/d_{yz}$ orbitals. (a) PDOS calculated using the low temperature crystal structure and G-AFM magnetic structure but with the on-site Coulomb interaction U = 0. (b) PDOS calculated using the low temperature crystal structure and G-AFM magnetic structure and with U = 2 eV. (c) PDOS calculated using high temperature crystal

structure and AFM-a type magnetic structure, a state similar to the field-induced one above the critical field, and with U = 2 eV.

**Figure 1**

M. Zhu et al,

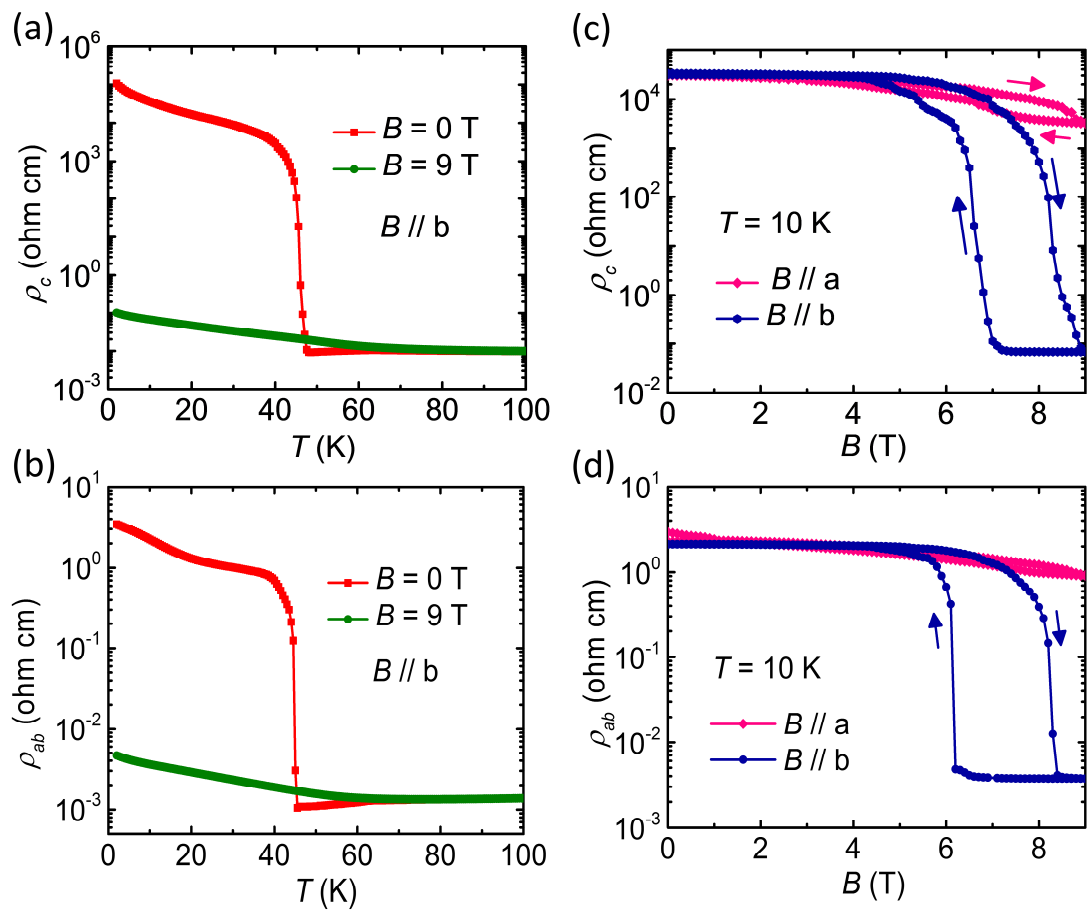

**Figure 2**

M. Zhu et al,

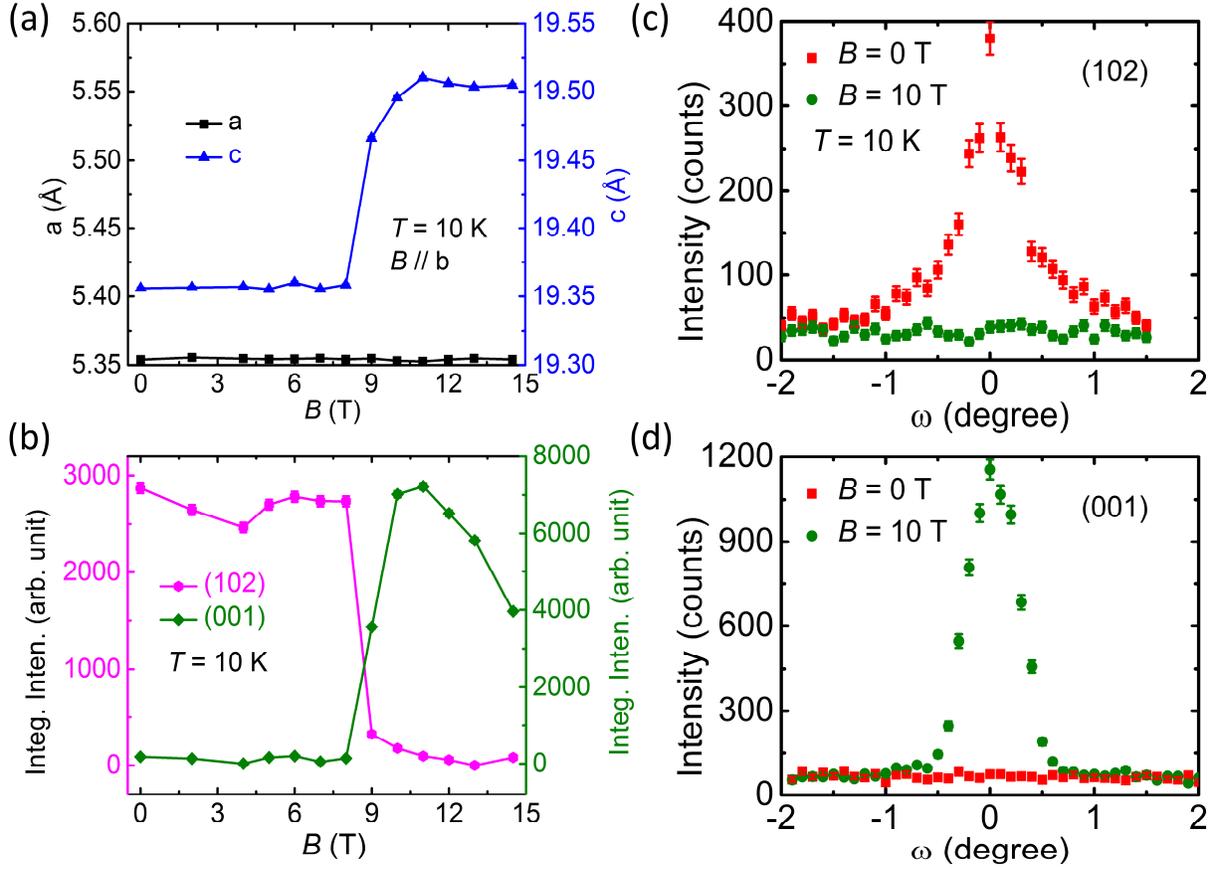

**Figure 3**

M. Zhu et al,

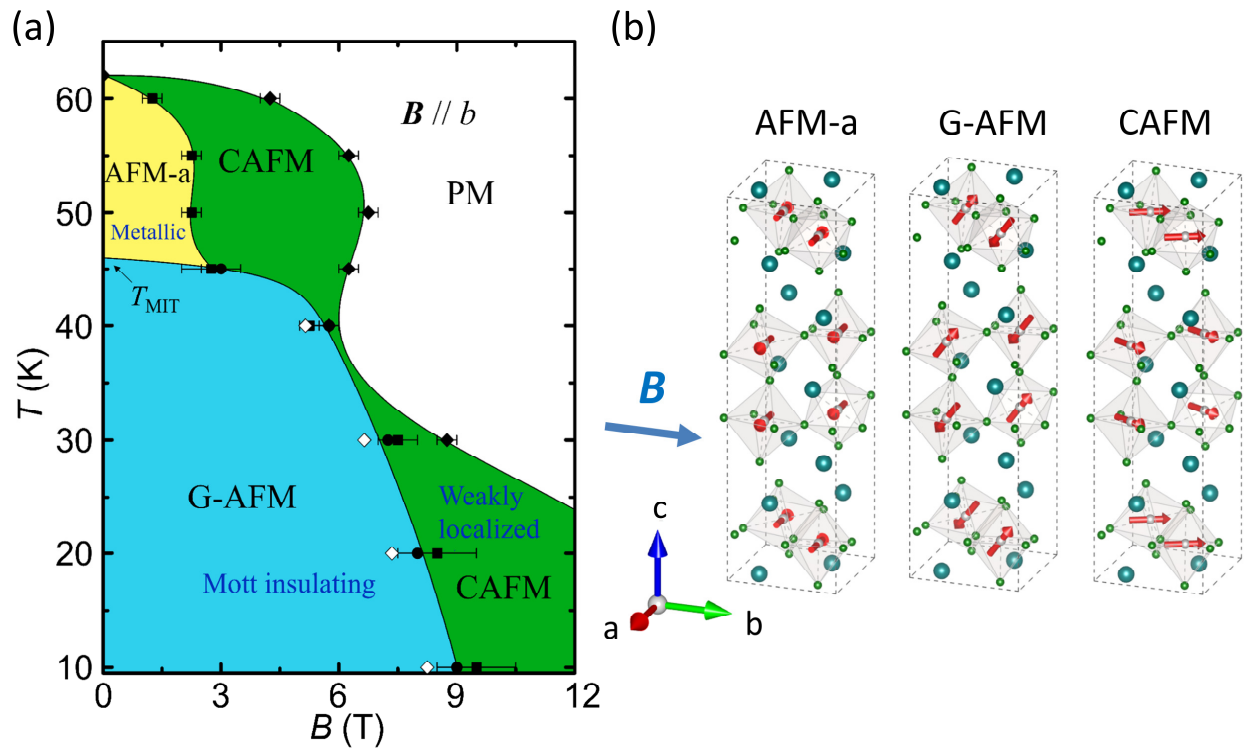

**Figure 4**

M. Zhu et al,

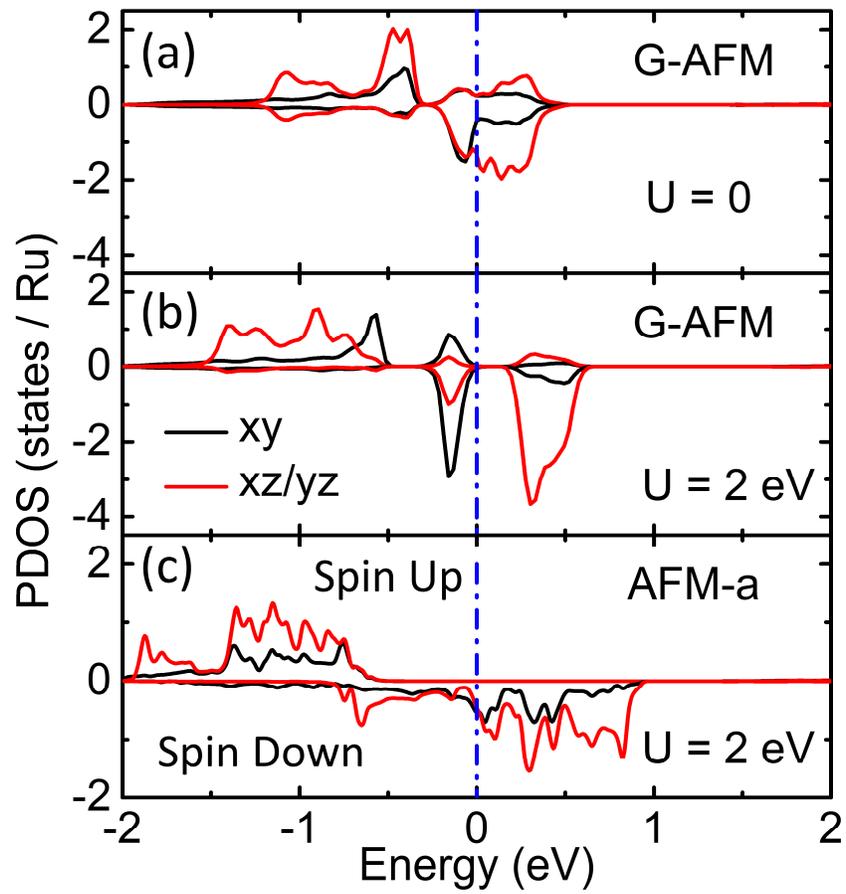